\documentclass{PoS}

\newcommand{\be}{\begin{eqnarray}}
\newcommand{\ee}{\end{eqnarray}}

\title{Basis light-front quantization approach to nucleon}

\ShortTitle{Basis light-front quantization approach to nucleon}

\author{\speaker{Chandan Mondal}$^{1,2,a}$, Siqi Xu$^{1,2,b}$, Jiangshan Lan$^{1,2,3,c}$, Xingbo Zhao$^{1,2,d}$, Yang Li$^{4,e}$, Dipankar Chakrabarti$^{5,f}$, James P. Vary$^{4,g}$\\
        $^1$Institute of Modern Physics, Chinese Academy of Sciences, Lanzhou 730000, China\\
        $^2$School of Nuclear Science and Technology, University of Chinese Academy of Sciences, \\Beijing 100049, China\\
        $^3$Lanzhou University, Lanzhou 730000, China\\
        $^4$Department of Physics and Astronomy, Iowa State University, Ames, IA 50011, USA\\
        $^5$Department of Physics, Indian Institute of Technology Kanpur, Kanpur 208016, India\\
        E-mail: $^a$\email{mondal@impcas.ac.cn}, $^b$\email{xsq234@impcas.ac.cn}, $^c$\email{jiangshanlan@impcas.ac.cn}, $^d$\email{xbzhao@impcas.ac.cn}, $^e$\email{leeyoung@iastate.edu}, $^f$\email{dipankar@iitk.ac.in}, 
        $^g$\email{jvary@iastate.edu}
}

\abstract{We obtain the light-front wavefunctions for the nucleon in the valence quark Fock space from an effective Hamiltonian, which includes the transverse and longitudinal confinement and the one-gluon exchange interaction with fixed coupling.
The wavefunctions are generated by solving the eigenvalue equation in a basis light-front quantization. Fitting the model parameters, the wavefunctions lead to good simultaneous description of electromagnetic form factors, radii, and parton distribution functions for the proton.
}

\FullConference{Light Cone 2019 - QCD on the light cone: from hadrons to heavy ions - LC2019\\
		16-20 September 2019\\
		Ecole Polytechnique, Palaiseau, France}

\begin{document}

\section{Introduction}

In recent years, Basis light front quantization (BLFQ) has emerged as one of the most
promising nonperturbative approaches which has been developed for solving many-body bound state problems in quantum field theories~\cite{Vary:2009gt,Zhao:2014xaa,Wiecki:2014,Li:2015zda}. It is based on Hamiltonian formalism and  incorporates the advantages of the light front dynamics \cite{Brodsky:1997de,Hiller:2016itl}. This formalism has been successfully applied to QED systems including the electron anomalous magnetic moment~\cite{Zhao:2014xaa} and the strong coupling bound-state positronium problem \cite{Wiecki:2014}. It has also been applied to heavy quarkonia~\cite{Li:2015zda,Li:2017mlw,Lan:2019img}, heavy-light mesons~\cite{Tang:2018myz}, light mesons~\cite{Jia:2018ary,Lan:2019vui,Lan:2019rba}, and proton \cite{Xu:2019xhk} as QCD bound states. 
Here, we employ an effective Hamiltonian that includes the holographic QCD confinement potential~\cite{Brodsky:2014yha} supplemented by the longitudinal confinement~\cite{Li:2015zda,Li:2017mlw} along with the one-gluon exchange (OGE) interaction with fixed coupling constant~\cite{Li:2015zda} to account for the dynamical spin effects. By solving its mass eigenstates in BLFQ, we generate the light-front wavefunctions (LFWFs) for the nucleon in the valence quark Fock space.
Fitting the quark mass, confining strength, and coupling constants, which are the model parameters, we obtain high quality descriptions of the electromagnetic form factors (FFs), radius and the parton distribution functions (PDFs) for the proton.

\section{Effective Hamiltonian and nucleon wavefunctions}
The LFWFs are obtained as the eigenfunctions of the light-front eigenvalue equation:
$
H_{\mathrm{eff}}\vert \Psi\rangle=M^2\vert \Psi\rangle.
$
We write the LF effective Hamiltonian $H_{\rm eff}$ in the leading Fock representation as \cite{Xu:2019xhk}
\begin{eqnarray}
H_{\rm eff}&=&\sum_a \frac{{\vec p}_{\perp a}^2+m_{a}^2}{x_a}+\frac{1}{2}\sum_{a\ne b}\kappa^4 \Big[x_ax_b({ \vec r}_{\perp a}-{ \vec r}_{\perp b})^2
-\frac{\partial_{x_a}(x_a x_b\partial_{x_b})}{(m_{a}+m_{b})^2}\Big]
\nonumber\\&&+\frac{1}{2}\sum_{a\ne b} \frac{C_F 4\pi \alpha_s}{Q^2_{ab}} \bar{u}_{s'_a}(k'_a)\gamma^\mu{u}_{s_a}(k_a)\bar{u}_{s'_b}(k'_b)\gamma^\nu{u}_{s_b}(k_b)g_{\mu\nu},
\end{eqnarray}\label{hami}
where $\sum_a x_a=1$, and $\sum_a {\vec p}_{\perp a}=0$. $m_{a/b}$ is the quark mass, and $\kappa$ is the confining strength. $x_a$ represents the LF momentum fraction carried by quark $a$. Meanwhile, $\vec{p}_\perp$ is the relative transverse momentum, while $\vec{r}_\perp={ \vec r}_{\perp a}-{ \vec r}_{\perp b}$, related to the holographic variable~\cite{Brodsky:2014yha}, is the transverse separation between two quarks. The last term in Eq. (\ref{hami}) represents the OGE interaction~\cite{Li:2015zda}

Following BLFQ, we expand nucleon state in terms of the two dimensional harmonic oscillator (`2D-HO') basis in the transverse direction and the discretized plane-wave basis in the longitudinal direction \cite{Vary:2009gt,Zhao:2014xaa}. Each single-quark basis state is identified using four quantum numbers, $\bar \alpha = \{k,n,m,\lambda\}$.
For a given quark $i$, the longitudinal momentum fraction $x$ is defined as
$
x_i=p_i^+/P^+=k_i/K,
$ where $K=\sum_i k_i$. We adopt antiperiodic boundary conditions so $k_i$ are positive half-odd integers.
The quantum numbers, $n$ and $m$, denote radial excitation and angular momentum projection, respectively, of the particle within the 2D-HO basis, $\phi_{nm}(\vec{p}_\perp)$ \cite{Vary:2009gt,Zhao:2014xaa}. 
The 2D-HO basis should form an efficient basis for systems subject to QCD confinement. For the quark spin, $\lambda$ is used to label the helicity. Our multi-body basis states have fixed values of the total angular momentum projection
$
M_J=\sum_i\left(m_i+\lambda_i\right).
$

The valence wavefunction in momentum space is then given by \cite{Xu:2019xhk}
\begin{eqnarray}\Psi^{M_J}_{\{x_i,\vec{p}_{\perp i},\lambda_i\}}&=&\sum_{\{n_im_i\}}\Bigg\{\psi(\{\bar{\alpha}_i\})\prod_{i=1}^{3}\frac{1}{b}\left(\frac{|\vec{p}_{\perp i}|}{b}\right)^{|m_i|}\nonumber\\
&\times &\sqrt{\frac{4\pi\times n_i!}{(n_i+|m_i|)!}}e^{i m_i\theta_i} L_{n_i}^{|m_i|}\left(-\frac{\vec{p}_{\perp i}^2}{b^2}\right)\exp{\left(-\frac{\vec{p}_{\perp i}^2}{2b^2}\right)}\Bigg\},\label{eq:psi_rs_basis_expansions}
\end{eqnarray}
where $\psi(\{\bar{\alpha}_i\})$ is the LFWF in the BLFQ basis obtained by diagnalizing Eq. (\ref{hami}) numerically. $b=0.6$ GeV is the HO scale parameter and $\tan(\theta)=p_2/p_1$. Here $L_n^{|m|}$ is the associated Laguerre function.
We truncate the infinite basis by introducing the longitudinal regulator $K_{\rm max}$ such that, $\sum_i k_i = K_{\rm max}$. In the transverse direction, we also truncate by limiting $N_\alpha=\sum_i (2n_i+| m_i |+1)$ for multi-particle basis state to $N_\alpha \le N_{\rm{max}}$. 
The basis truncation corresponds to a UV regulator $\Lambda_{\rm UV}\sim b\sqrt{N_{\rm max}}$. Since we are modeling the proton at a low-resolution scale, we select $N_{\rm max}=10$ and $K_{\rm max}=16.5$. To attempt to simulate the effect of higher Fock spaces and the other QCD interactions,  we use a different quark mass in the kinetic energy, $m_{\rm q/KE}$ and the OGE interaction, $m_{\rm q/OGE}$. We set our parameters $\{m_{\rm q/KE},~m_{\rm q/OGE},~\kappa,~\alpha_s\}=\{0.3~{\rm GeV},~0.2~{\rm GeV},~0.34~{\rm GeV},~1.1\}$ to fit the proton mass and the flavor Dirac FFs \cite{Cates:2011pz,Qattan:2012zf,Diehl:2013xca,Mondal:2015uha,Mondal:2016xpk}. For numerical convenience, we use a small gluon mass regulator ($\mu_g=0.05$ GeV) in the OGE interaction. We find that our results are insensitive to $0.08>\mu_g>0.01$ GeV. 
\section{Form Factors and PDFs of the nucleon}

In the LF formalism, the flavor Dirac $F^q_1(Q^2)$ and Pauli $F^q_2(Q^2)$ FFs in the proton can be expressed in terms of overlap integrals as \cite{BDH}
\begin{eqnarray}
F_1^q(Q^2)= 
\int_D \Psi^{\uparrow*}_{\{x^{\prime}_i,
\vec{p}^{\prime}_{\perp i},\lambda_i\}} \Psi^{\uparrow}_{\{x_i,
\vec{p}_{\perp i},\lambda_i\}}, \quad
F_2^q(Q^2)= -\frac{2 M }{(q^1-iq^2)}
\int_D \Psi^{\uparrow*}_{\{x^{\prime}_i,
\vec{p}^{\prime}_{\perp i},\lambda_i\}} \Psi^{\downarrow}_{\{x_i,
\vec{p}_{\perp i},\lambda_i\}},
\end{eqnarray}
with
$\int_D\equiv \sum_{\lambda_i}\int \prod_i[{dx d^2
\vec{p}_{\perp}\over 16 \pi^3}]_i 16 \pi^3 \delta\left(1-\sum x_j\right)
\delta^2\left(\sum \vec{p}_{\perp j}\right).
$
For the struck quark of flavor $q$, 
 ${x'}_1=x_1; 
~{\vec{p}'}_{\perp 1}=\vec{p}_{\perp 1}+(1-x_1) \vec{q}_\perp$ and  ${x'}_i={x_i}; ~{\vec{p}'}_{\perp i}=\vec{p}_{\perp i}-{x_i} \vec{q}_\perp$ for  the spectators ($i=2,3$). We consider the frame where the momentum transfer $q=(0,0,\vec{q}_{\perp})$, thus $Q^2=-q^2=\vec{q}_{\perp}^2$.
The nucleon Sachs form factors are written in the terms of Dirac and Pauli form factors,
\begin{eqnarray}
G_E^N(Q^2)= F_1^{N}(Q^2) - \frac{Q^2}{4M_N^2} F_2^{N}(Q^2), \quad\quad
G_M^N(Q^2)= F_1^{N}(Q^2) + F_2^N(Q^2),
\end{eqnarray}
where $F_{1/2}^N=\sum_q e_q F_{1/2}^{q/N}$ is the Dirac (Pauli) form factors of the nucleon. The electromagnetic radii of the nucleon can be obtained from
\begin{eqnarray}
{\langle r^2_{E}\rangle }^N=-6 \frac{d G_{E}^N(Q^2)}{dQ^2}{\Big\vert}_{Q^2=0}, \quad\quad
{\langle r^2_{M}\rangle}^N=-\frac{6}{G_{M}^N(0)} \frac{d G_{M}^N(Q^2)}{dQ^2}{\Big\vert}_{Q^2=0}.
\end{eqnarray}

\begin{figure}[htp]
\begin{center}
\includegraphics[width=0.4\textwidth]{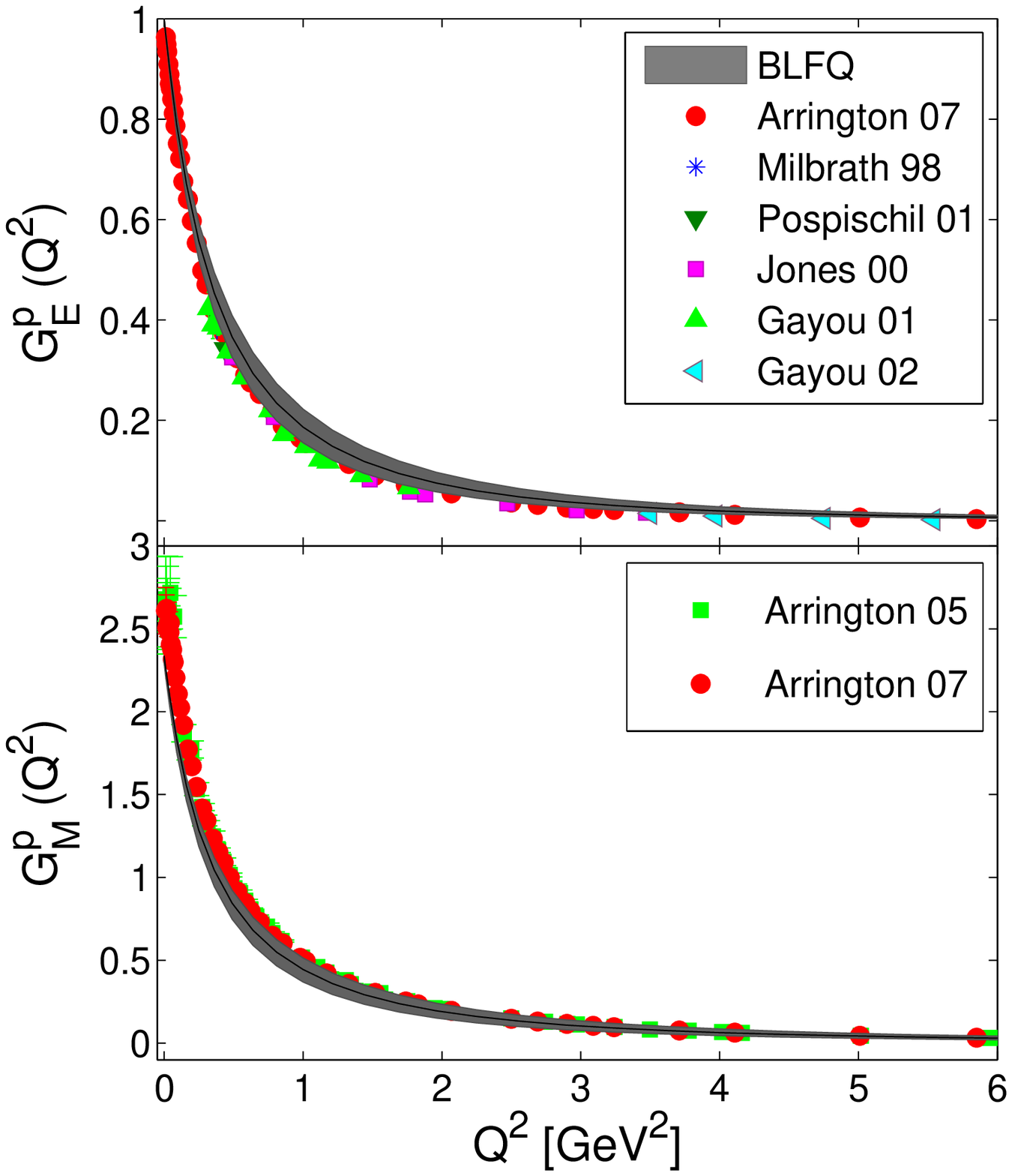}
\includegraphics[width=0.4\textwidth]{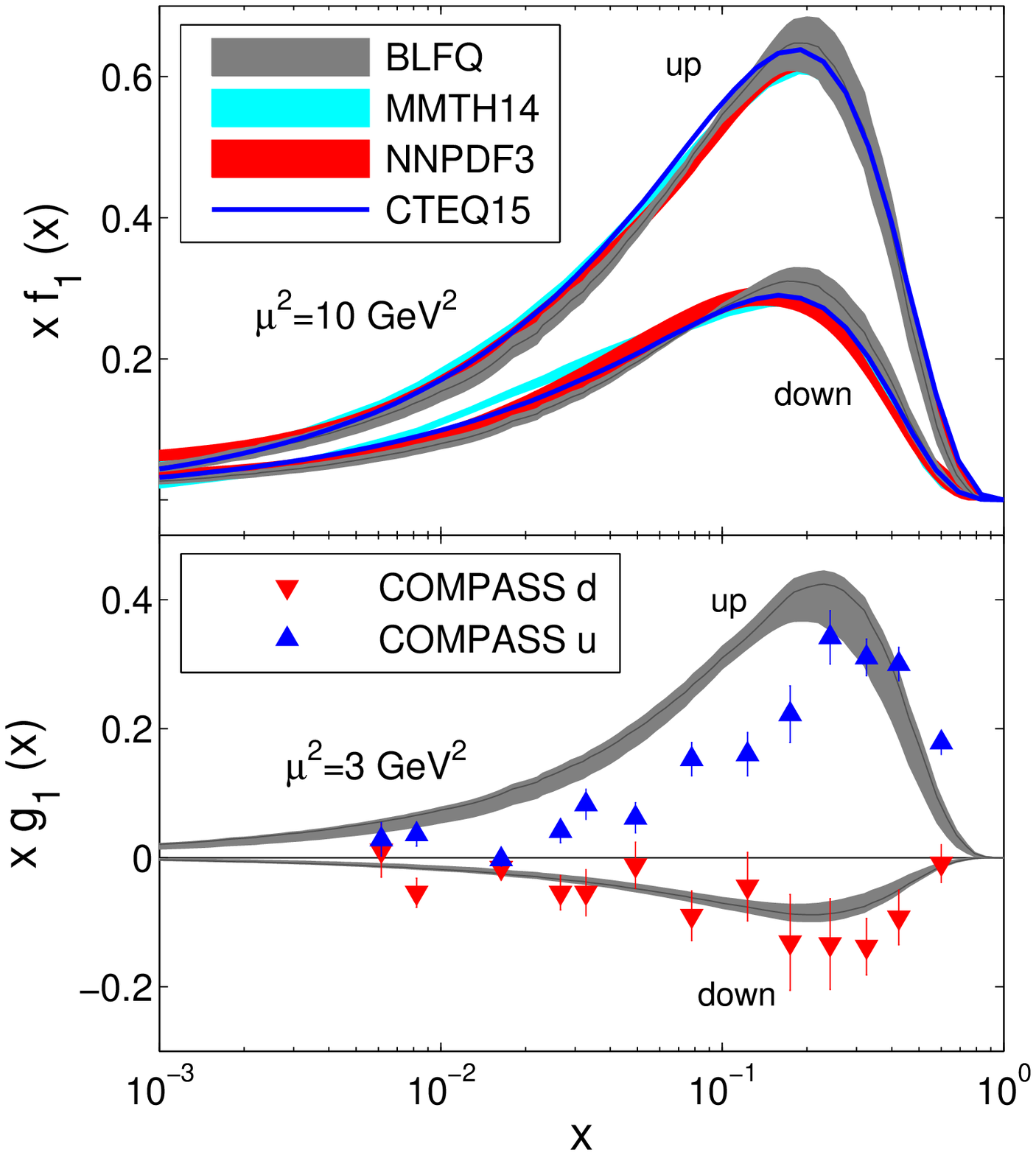}
\caption{Left panel: proton Sach's FFs $G_E^p (Q^2)$ and  $G_M^p (Q^2)$
as functions of $Q^2$. 
Right panel: comparison for $xf_1(x)$ in the proton from BLFQ (gray
bands) and global fits; and comparison for $xg_1(x)$ from BLFQ (gray
bands) and measured data from COMPASS. The experimental data can be found in Ref~\cite{Xu:2019xhk,Chakrabarti:2013dda}.
}
\label{figGEM}
\end{center}
\end{figure}
In the left panel of Fig. \ref{figGEM}, we show the $Q^2$ dependence of the proton electric and the magnetic Sach's FFs. 
Overall, we  obtain a reasonable agreement between theory and experiment for the proton electric FFs. At large $Q^2$, the magnetic form factor is also in good agreement with the data. However, our magnetic form factor at low $Q^2$ exhibits a small deviation from the data. It should be noted that the neglected higher Fock components $|qqqq\bar{q}\rangle$ can have a significant effect on the magnetic form factor.
\begin{table}[htp]	
\centering
\caption{The nucleon radii in BLFQ are compared with the experimental data~\cite{Tanabashi:2018oca} and lattice results~\cite{Alexandrou:2018sjm}.}
\label{tab:radii}
\begin{tabular}{cccccc}
\hline\hline
Quantity  	   					~& BLFQ	          			~& Measurement       			~& Lattice   		\\
\hline
$r^{\rm{P}}_E ~\rm{fm}$    		~& $0.802^{+0.042}_{-0.040}$	~& $0.833\pm 0.010$		  	~& $0.742(13)$		\\
$r^{\rm{P}}_M ~\rm{fm}$    		~& $0.834^{+0.029}_{-0.029}$	~& $0.851\pm 0.026$  		~& $0.710(26)$		\\
\hline
$(r^{\rm{N}}_E)^2 ~\rm{fm}^2$   ~& $-0.033\pm 0.198$	  		~& $-0.1161\pm 0.0022$ 	  	~& $-0.074(16)$		\\
$r^{\rm{N}}_M ~\rm{fm}$    		~& $0.861^{+0.021}_{-0.019}$	~& $0.864^{+0.009}_{-0.008}$ ~& $0.716(29)$		\\
\hline\hline
\end{tabular}
\end{table}
We present our computed radii in Table \ref{tab:radii} and compare with measured data~\cite{Tanabashi:2018oca} as well as with recent lattice QCD calculations~\cite{Alexandrou:2018sjm}. Here again,
except for the charge radius of the neutron, we find reasonable agreement with experiment.

With our LFWFs, the proton's valence quark PDFs at leading twist are given by
\begin{eqnarray}
f_1^q=
\int_D \Psi^{\uparrow*}_{\{x^\prime_i,
\vec{p}^\prime_{\perp i},\lambda_i\}} \Psi^{\uparrow}_{\{x_i,
\vec{p}_{\perp i},\lambda_i\}} \delta(x-x_1), \nonumber
\quad
g_1^q=
\int_D (\Lambda)~\Psi^{\uparrow*}_{\{x^\prime_i,
\vec{p}^\prime_{\perp i},\lambda_i\}} \Psi^{\uparrow}_{\{x_i,
\vec{p}_{\perp i},\lambda_i\}}\delta(x-x_1) , 
\end{eqnarray}
where $\Lambda=1(-1)$ depends on the struck quark helicity $\lambda_1=\frac{1}{2}(-\frac{1}{2})$.
 At the model scale relevant to constituent quark masses which are several hundred $\mathrm{MeV}$, the unpolarized PDFs for the valence quarks are normalized as
$
\int_{0}^{1}f_1^u(x)\,dx =2, \int_{0}^{1}f_1^d(x)\,dx=1.
$
We also have the following momentum sum rule: 
$
\int_{0}^{1}x\,f_1^u(x)\,dx +\int_{0}^{1}x\,f_1^d(x)\,dx=1. 
$

The right panel of Fig. \ref{figGEM} shows our results for the valence quark unpolarized and spin dependent PDFs of the proton, where we compare the valence quark distribution after QCD evolution with the  global fits by MMHT14, NNPDF3.0, and CTEQ15 Collaborations. The error bands in our evolved distributions are due to the spread in the initial scale $\mu_{0}^2=0.195\pm 0.020$ GeV$^2$ and the uncertainties in the coupling constant, $\alpha_s=1.1\pm0.1$. We determine ${\mu_{0}^2=0.195\pm0.020~\rm{GeV}^2}$ by requiring the result after QCD evolution to produce the total first moments of the valence quark unpolarized PDFs from the global data fits with average values, $\langle x\rangle_{u_v+d_v}=0.37\pm 0.01$ at $\mu^2=10$ GeV$^2$. Our unpolarized valence PDFs for both up and down quarks are found to be in good agreement with the global fits. Meanwhile, we evolve the spin dependent PDFs from our model scale to the relevant experimental scale $\mu^2=3$ GeV$^2$ and find that the down quark helicity PDF agrees well with measured data from COMPASS Collaboration. However, for the up quark, our helicity PDF tends to overestimate the data below $x\sim 0.3$.
\\
\\
{\it Acknowledgment:}
CM is supported by the National Natural Science Foundation of China (NSFC) under the Grant Nos. 11850410436 and 11950410753. XZ is supported by  Key Research Program of Frontier Sciences, CAS, Grant No ZDBS-LY-7020. JPV is supported by the Department of Energy under Grants No. DE-FG02-87ER40371, and No. DE-SC0018223 (SciDAC4/NUCLEI). A portion of the computational resources were provided by the National Energy Research Scientific Computing Center (NERSC), which is supported by the Office of Science of the U.S. Department of Energy under Contract No.DE-AC02-05CH11231.


\begin{thebibliography}{99}

  
\bibitem{Vary:2009gt}
  J.~P.~Vary {\it et al.},
  Phys.\ Rev.\ C {\bf 81}, 035205 (2010).
  
  \bibitem{Zhao:2014xaa}
  X.~Zhao, H.~Honkanen, P.~Maris, J.~P.~Vary and S.~J.~Brodsky,
  Phys.\ Lett.\ B {\bf 737}, 65 (2014).

\bibitem{Wiecki:2014}
  P.~Wiecki, Y.~Li, X.~Zhao, P.~Maris and J.~P.~Vary,
  Phys.\ Rev.\ D {\bf 91}, no. 10, 105009 (2015).
  
\bibitem{Li:2015zda}
  Y.~Li, P.~Maris, X.~Zhao and J.~P.~Vary,
  Phys.\ Lett.\ B (2016).
  
   
  
  \bibitem{Brodsky:1997de}
  S.~J.~Brodsky, H.~C.~Pauli and S.~S.~Pinsky,
  Phys.\ Rept.\  {\bf 301}, 299 (1998). 
  
  \bibitem{Hiller:2016itl} 
  J.~R.~Hiller,
  Prog.\ Part.\ Nucl.\ Phys.\  {\bf 90}, 75 (2016).
  
\bibitem{Li:2017mlw} 
  Y.~Li, P.~Maris and J.~P.~Vary,
  Phys.\ Rev.\ D {\bf 96}, no. 1, 016022 (2017).
  
  \bibitem{Lan:2019img} 
  J.~Lan, C.~Mondal, M.~Li, Y.~Li, S.~Tang, X.~Zhao and J.~P.~Vary,
  arXiv:1911.11676 [nucl-th].
  
  
  
\bibitem{Tang:2018myz} 
  S.~Tang, Y.~Li, P.~Maris and J.~P.~Vary,
  Phys.\ Rev.\ D {\bf 98}, no. 11, 114038 (2018).
  
  
\bibitem{Jia:2018ary} 
  S.~Jia and J.~P.~Vary,
  Phys.\ Rev.\ C {\bf 99}, no. 3, 035206 (2019).
  
\bibitem{Lan:2019vui} 
  J.~Lan, C.~Mondal, S.~Jia, X.~Zhao and J.~P.~Vary,
  Phys.\ Rev.\ Lett.\  {\bf 122}, no. 17, 172001 (2019).
  
  \bibitem{Lan:2019rba} 
  J.~Lan, C.~Mondal, S.~Jia, X.~Zhao and J.~P.~Vary,
  arXiv:1907.01509 [nucl-th].
  
  \bibitem{Xu:2019xhk} 
  C.~Mondal, S.~Xu, J.~Lan, X.~Zhao, Y.~Li, D.~Chakrabarti and J.~P.~Vary,
  arXiv:1911.10913 [hep-ph].
  
    \bibitem{Brodsky:2014yha}
  S.~J.~Brodsky, G.~F.~de Teramond, H.~G.~Dosch and J.~Erlich,
  Phys.\ Rept.\  {\bf 584}, 1 (2015).
  
\bibitem{Cates:2011pz} 
  G.~D.~Cates, C.~W.~de Jager, S.~Riordan and B.~Wojtsekhowski,
  Phys.\ Rev.\ Lett.\  {\bf 106}, 252003 (2011).

\bibitem{Qattan:2012zf} 
  I.~A.~Qattan and J.~Arrington,
  Phys.\ Rev.\ C {\bf 86}, 065210 (2012).
  
  \bibitem{Diehl:2013xca} 
  M.~Diehl and P.~Kroll,
  Eur.\ Phys.\ J.\ C {\bf 73}, no. 4, 2397 (2013).
  
  \bibitem{Mondal:2015uha} 
  C.~Mondal and D.~Chakrabarti,
  Eur.\ Phys.\ J.\ C {\bf 75}, no. 6, 261 (2015).
  
  \bibitem{Mondal:2016xpk} 
  C.~Mondal,
  Phys.\ Rev.\ D {\bf 94}, no. 7, 073001 (2016).


  \bibitem{BDH}  S. J. Brodsky, M. Diehl, D. S. Hwang, Nucl. Phys. B {\bf 596}, 99 (2001).
  
  \bibitem{Tanabashi:2018oca} 
  M.~Tanabashi {\it et al.} [Particle Data Group],
  Phys.\ Rev.\ D {\bf 98}, no. 3, 030001 (2018).
  
  \bibitem{Alexandrou:2018sjm} 
  C.~Alexandrou, $et.~al.$,
  Phys.\ Rev.\ D {\bf 100}, no. 1, 014509 (2019).
  
  \bibitem{Chakrabarti:2013dda} 
  D.~Chakrabarti and C.~Mondal,
  Eur.\ Phys.\ J.\ C {\bf 73}, 2671 (2013).
  
  


\end{thebibliography}
\end{document}